\documentclass[11pt]{article}
\usepackage[utf8]{inputenc}
\usepackage[T1]{fontenc}
\usepackage{listings}
\usepackage{xcolor}

\lstdefinelanguage{yaml}{
  keywords={true,false,null,y,n},
  keywordstyle=\color{darkgray}\bfseries,
  basicstyle=\footnotesize\ttfamily,
  sensitive=false,
  comment=[l]{\#},
  morecomment=[s]{/*}{*/},
  commentstyle=\color{purple}\ttfamily,
  stringstyle=\color{blue}\ttfamily,
  moredelim=[l][\color{orange}]{\&},
  moredelim=[l][\color{magenta}]{*},
  moredelim=**[il][\color{red}]{:},
  morestring=[b]',
  morestring=[b]"
}
\usepackage{graphicx}
\usepackage{amsmath}
\usepackage{amssymb}
\usepackage{enumitem}
\usepackage{geometry}
\usepackage{cite}
\usepackage[hidelinks]{hyperref}
\usepackage{orcidlink}

\title{ADA: Automated Moving Target Defense for AI Workloads via Ephemeral Infrastructure-Native Rotation in Kubernetes}
\author{
    Akram Sheriff\,\orcidlink{0009-0004-4511-6637} \\
    \texttt{isheriff@cisco.com} \\
    Cisco Systems \\
    \and
    Ken Huang\,\orcidlink{0009-0004-6502-3673} \\
    \texttt{ken.huang@distributedapps.ai} \\
    Distributedapps.ai \\
    \and
    Zsolt Nemeth \\
    \texttt{zsolt@r6security.com} \\
    R6 Security \\
    \and
    Madjid Nakhjiri \\
    \texttt{madjid@avencezgroup.com} \\
    Independent Security Researcher \\
}
\date{\today}

\begin{document}

\maketitle

\begin{abstract}
This paper introduces the Adaptive Defense Agent (ADA), an innovative Automated Moving Target Defense (AMTD) system designed to fundamentally enhance the security posture of AI workloads. ADA operates by continuously and automatically rotating these workloads at the infrastructure level, leveraging the inherent ephemerality of Kubernetes pods. This constant managed churn systematically invalidates attacker assumptions and disrupts potential kill chains by regularly destroying and respawning AI service instances. This methodology, applying principles of chaos engineering as a continuous, proactive defense, offers a paradigm shift from traditional static defenses that rely on complex and expensive confidential or trusted computing solutions to secure the underlying compute platforms, while at the same time agnostically supporting the latest advancements in agentic and nonagentic AI ecosystems and solutions such as agent-to-agent (A2A) communication frameworks or model context protocols (MCP). This AI-native infrastructure design, relying on the widely proliferated cloud-native Kubernetes technologies, facilitates easier deployment, simplifies maintenance through an inherent zero trust posture achieved by rotation, and promotes faster adoption. We posit that ADA's novel approach to AMTD provides a more robust, agile, and operationally efficient zero-trust model for AI services, achieving security through proactive environmental manipulation rather than reactive patching.
\end{abstract}


\section{Introduction}
\textbf{T}he speed of innovation in Artificial Intelligence (AI) is revolutionizing industries, with AI, Agentic AI workloads increasingly deployed as microservices, such as NVIDIA NIM (NVIDIA Inference Microservices) \cite{nvidia_nim}. These services are predominantly hosted on container orchestration platforms like Kubernetes, which offer scalability and resilience, but also introduce new security challenges \cite{kubernetes_security}.

Current security paradigms, often reliant on static defenses, struggle to adequately protect these fluid environments. One can go to great lengths in server and device design to achieve confidential or trusted computing for container images. Traditionally, sensitive applications such as those in financial, critical infrastructure, and defense have relied on expensive, complicated hardware-based secure execution environments such as ARM TrustZone, TCG TPM or similar. These solutions require not only special hardware considerations but also very careful design and implementation of hardware-firmware interactions to keep the workload execution within a trusted boundary and its interactions with the outside world secure.

The rise of agentic AI applications, in particular those deploying multiple agents, and leveraging LLM models with intensive interactions with external context through prompts, MCP Servers, or tool interactions would make the design and implementation of a confidential computing environment even more complex. However, microservices systems are always one vulnerable misconfiguration or container image/firmware image vulnerability away from an exploit even in hardware-secured environments. As we will see in the MAESTRO threat modeling section of this paper, an AI system has many more components than a traditional application system that can be exploited. Models can be modified to include malware that perform privilege escalations and gain access to critical system components, input prompts into the system can include malware, and access to various AI system artifacts can be compromised. Attackers can exploit the predictability of static configurations to establish persistence and execute advanced attacks. Although NVIDIA provides tools and guidance to secure NIMs \cite{nvidia_nim}, the security of the underlying infrastructure remains a critical concern. This paper argues for a transformative approach to securing these AI workloads by making the infrastructure itself an active defense mechanism.

We introduce Adaptive Defense Agent (ADA), a system rooted in the principles of Automated Moving Target Defense (AMTD) \cite{mtd_survey}. ADA innovatively applies AMTD by continuously rotating AI workload instances (Kubernetes pods) at the infrastructure level. This automated ``destroy and respawn'' cycle makes the AI service instances ephemeral, directly invalidating attacker assumptions about the target's stability and accessibility. This concept draws parallels with chaos engineering, where controlled disruption is used not just for resilience testing, but as a continuous, proactive security strategy \cite{chaos_engineering}.

The key innovation of ADA lies in its infrastructure-native design, which leverages many Kubernetes components and features, which contrasts sharply with the complexities often introduced by agent-to-agent (A2A) security models. A2A solutions can require new security controls, complex identity management, and substantial maintenance efforts. ADA bypasses these by leveraging existing Kubernetes capabilities, offering:

\begin{itemize}
    \item \textbf{Simplified Deployment:} Integrating as a Kubernetes controller with minimal changes to existing deployments.
    \item \textbf{Streamlined Maintenance:} Achieving a zero-trust posture through rotation, reducing the need for complex instance-level trust management.
    \item \textbf{Accelerated Adoption:} Aligning with the operational practices of the DevOps, LLMops and MLOps teams familiar with Kubernetes \cite{kubernetes_security}.
\end{itemize}

ADA proposes a zero trust model ``by design''---where the fleeting nature of an instance minimizes its implicit trust---rather than ``by patching.'' This paper details the architecture of ADA, its operational advantages, and its potential to significantly enhance the security of modern AI deployments by providing a proactive, agile, and resilient defense.

\section{Related Work}
ADA builds on established concepts in MTD, chaos engineering, and Kubernetes security, synthesizing them into a novel defense strategy for AI workloads.

\subsection{Moving Target Defense (MTD)}
MTD aims to create dynamic, unpredictable systems to increase attacker costs and reduce their window of opportunity \cite{mtd_survey}. The techniques span the network, platform, application, and data layers. AMTD focuses on automating these defenses, crucial for cloud environments. ADA is an AMTD system that focuses on infrastructure-level rotation (re-puffing) for AI workloads. Academic research often highlights strategies such as spatial, temporal, and configuration randomization to thwart attackers.

\subsection{Chaos Engineering for Security}
Chaos engineering involves controlled experiments to build confidence in the resilience of a system \cite{chaos_engineering}. Applying this to security involves proactive disruption to uncover weaknesses or make the environment unpredictable for attackers. ADA operationalizes this as a continuous defensive measure, where the ``chaos'' of pod rotation is the intended security mechanism.

\subsection{Kubernetes Security}
Kubernetes provides security features such as RBAC, pod security standards, and network policies \cite{kubernetes_security}. Zero-trust principles are often implemented via service meshes. ADA complements these by addressing the temporal aspect of instance security, assuming a breach is possible, and aiming to limit its lifespan through rotation. Secure deployment on Kubernetes platforms such as Amazon EKS also involves adherence to the best practices outlined by cloud providers \cite{aws_eks}.

\subsection{AI Workload Security (NVIDIA NIMs Focus)}
Securing AI models, such as NIM, involves protecting weights, data, and the inference process \cite{nvidia_nim}. NVIDIA offers guidance and tools for this. ADA adds a crucial infrastructure security layer by making NIM instances ephemeral, hindering attackers trying to establish persistence on these inference servers.

\subsection{Ephemeral Infrastructure}
The security benefits of ephemeral (short-lived) and immutable infrastructure are well recognized. Replacing instances rather than patching them reduces configuration drift and eliminates persistent threats. ADA embodies this for AI workloads as a proactive AMTD strategy.

\subsection{Context-Aware Resource Mutation}
Kubernetes manifests for agent deployments are dynamically mutated using CRD-defined policies. Mutation parameters are derived from telemetry feeds (e.g., Kube-state-metrics, OPA Gatekeeper alerts) and include GPU access changes, updated runtime environments, and new container images.

\subsubsection{ADA Mutation Engine or Adaptive Mutator Layer}
Beyond schedule-based rotation, ADA can be extended with context-sensitive Kubernetes manifest mutation, enabling deeper environmental adaptation. Mutation logic is defined through CRD-based policies and triggered by real-time telemetry sources ----such as Kube-state metrics, Prometheus alerts, or OPA Gatekeeper violations. These inputs drive selective updates to pod specifications, including:

\begin{itemize}
    \item Switching container images based on risk level or freshness
    \item Adjusting access to specialized resources (e.g., GPUs or ephemeral volumes)
    \item Modifying runtime environments (e.g., switching from Python to CUDA contexts)
\end{itemize}

This approach introduces an additional control layer that changes workload posture based on observed conditions, enhancing ADA's resilience, and aligning with the principles of self-healing, adaptive infrastructure.

\begin{lstlisting}[language=yaml,caption={Context Mutation Policy Example},label={lst:mutation_policy},basicstyle=\footnotesize\ttfamily,breaklines=true,frame=single]
apiVersion: ADA.security.r6.dev/v1
kind: ContextMutationPolicy
metadata:
  name: nim-risk-based-mutation
spec:
  selector:
    matchLabels:
      app: nim-inference
  triggers:
    telemetrySources:
      - type: PrometheusAlert
        name: high_gpu_usage
      - type: GatekeeperViolation
        constraint: disallowed-hostpath
  mutations:
    - type: ContainerImageUpdate
      containerName: nim
      newImage: "nvcr.io/nim/secure-nim:latest"
    - type: ResourceAdjustment
      containerName: nim
      resources:
        limits:
          nvidia.com/gpu: "0"
        requests:
          cpu: "500m"
          memory: "256Mi"
    - type: EnvPatch
      containerName: nim
      env:
        - name: RUNTIME_MODE
          value: "SECURE"
\end{lstlisting}

\subsubsection{YAML Context Mutation Policy Explanation}

The above ContextMutationPolicy demonstrates ADA's adaptive security capabilities through intelligent workload mutation. This YAML configuration implements the following security automation as below:

\textbf{Target Selection:}
\begin{itemize}
    \item Applies to any pod with label \texttt{app: nim-inference}
    \item Enables selective targeting of NVIDIA NIM inference workloads at runtime.
    \item Supports fine-grained policy application across heterogeneous AI deployments
\end{itemize}

\textbf{Trigger Conditions:}
\begin{itemize}
    \item \textbf{Prometheus Alert Monitor:} Watches for alert named \texttt{high\_gpu\_usage} indicating resource exhaustion or potential cryptomining, cryptojacking attacks
    \item \textbf{OPA Gatekeeper Integration:} Detects OPA gatekeeper policy violations for \texttt{disallowed-hostpath} constraint, preventing unauthorized filesystem access
    \item \textbf{Real-time Response:} Enables immediate reaction to security events without manual intervention
\end{itemize}

\textbf{Automated Mutation Actions:}
\begin{enumerate}
    \item \textbf{Secure Container Image Update:} Transitions to hardened image \texttt{nvcr.io/nim/secure-nim:latest} with enhanced security controls
    \item \textbf{Resource Privilege Revocation:} Revokes GPU access (\texttt{nvidia.com/gpu: "0"}) and enforces conservative CPU/memory limits to contain potential threats
    \item \textbf{Runtime Environment Modification:} Injects \texttt{RUNTIME\_MODE=SECURE} environment variable to switch inference execution to security-prioritized mode
\end{enumerate}

This policy exemplifies the ADA's philosophy of proactive defense through infrastructure manipulation, automatically degrading workload privileges when security anomalies are detected while maintaining service availability in a restricted operational mode.

\subsection{MCP Security}
Model Context Protocol (MCP) as a mechanism to provide agentic AI solutions to a user running an AI application that can deploy Gen AI models and a host of software tools through the services provided by a MCP server. MCP protocol has been subject to some scrutiny by the cybersecurity community. Many vulnerabilities have been identified, ranging from tool poisoning, MCP server poisoning, MCP client-MCP server communication vulnerabilities, etc. All components within the MCP ecosystem, ranging from the AI application to the MCP server and the underlying tools, can benefit from platform security hardening. ADA can be seen as very suitable in any deployment where any of the MCP components are running as container images.

\subsection{A2A Security}
Researchers in \cite{a2a_security} examine how to securely build AI applications using Google's Agent-to-Agent (A2A) protocol, which allows autonomous agents to communicate and collaborate. The A2A protocol uses machine-readable AgentCards for discovery and secure JSON-RPC requests for task execution, with real-time updates via server-sent events.

The authors identify security risks such as prompt injection, data poisoning, replay attacks, and network threats. To address these, they recommend digitally signing AgentCards, strict input validation, robust authentication, and fine-grained access controls.

Case studies include secure collaborative document editing and privacy-preserving data analysis. The paper concludes that while A2A is powerful in building secure agentic AI systems, its effectiveness relies on strong cryptographic safeguards, continuous threat assessment, and zero-trust security practices. Although \cite{a2a_security} proposes valid controls for A2A security, this paper presents a complementary approach that can serve as a final mitigation strategy when other security measures have been exhausted or compromised.

\section{The ADA System: An Innovative Defense}
ADA introduces a novel AMTD strategy by treating AI workload instances as inherently ephemeral components, continuously refreshed to maintain a secure baseline.

\subsection{Core Principles: A New Defensive Posture}
\begin{itemize}
    \item \textbf{AMTD via Continuous Rotation:} The cornerstone of ADA is the automated, continuous cycle of destroying and respawning AI workload instances (pods), driven by either a fixed schedule or anomaly detection. This dual-trigger approach ensures proactive disruption of attacker persistence, including threats that may evade detection.
    \item \textbf{Infrastructure-Native Simplicity:} ADA integrates directly with Kubernetes, minimizing new architectural components and aligning with cloud-native operations \cite{kubernetes_security}.
    \item \textbf{Zero-Trust through Ephemerality:} ADA actualizes a zero trust position at the instance level. An instance is never trusted for long; trust is vested in the orchestration of its constant, clean refresh.
    \item \textbf{Chaos as Proactive Defense:} The ``chaotic'' but controlled rotation is the defense itself, fostering an environment inherently hostile to attackers \cite{chaos_engineering}.
\end{itemize}

\subsection{Architecture: Lean and Integrated}
\begin{figure}[htbp]
\centering
\includegraphics[width=0.8\textwidth]{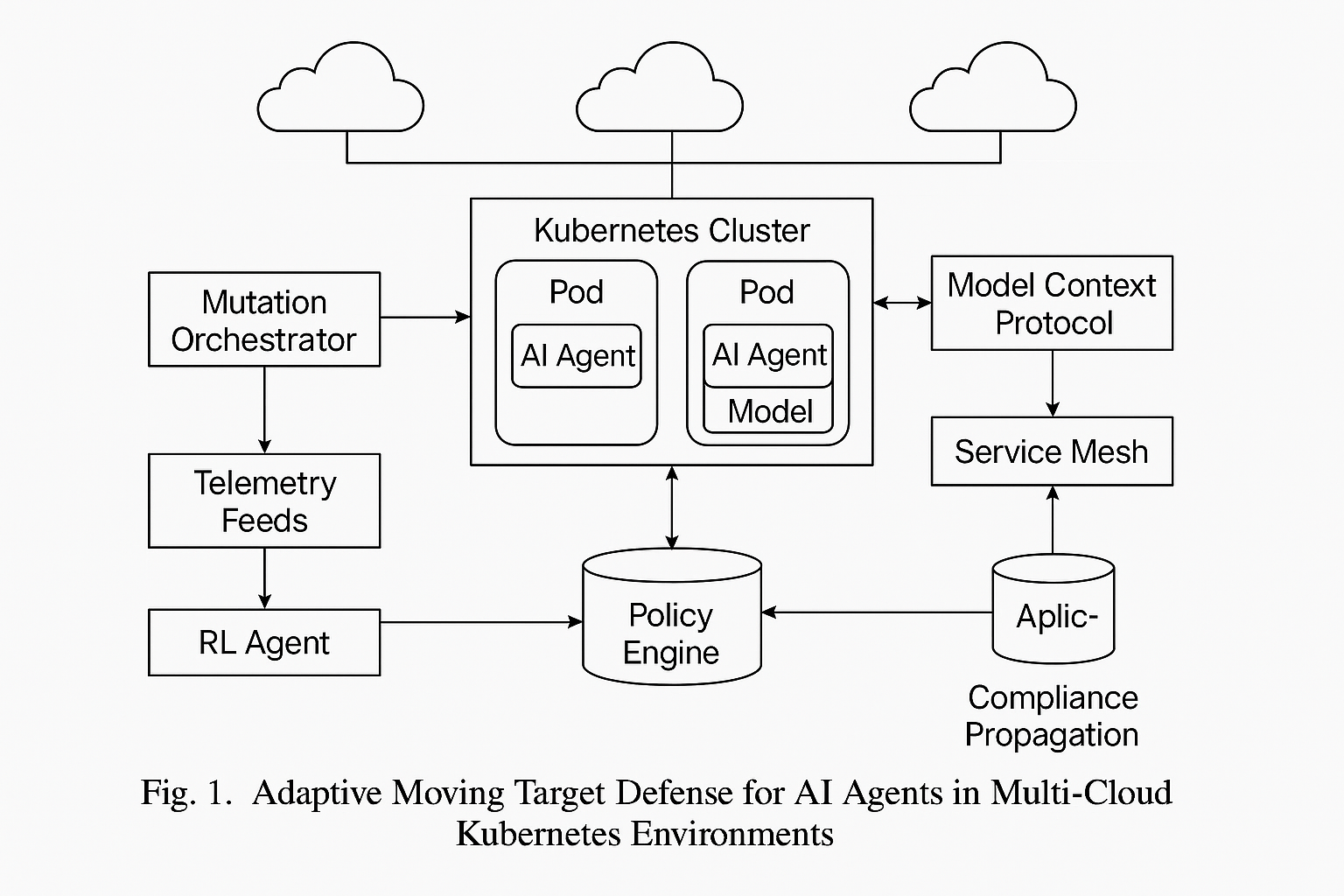}
\caption{ADA Architecture Diagram}
\label{fig:architecture}
\end{figure}

Key components:
\begin{itemize}
    \item \textbf{ADA Controller:} Orchestrates the rotation based on defined policies.
    \item \textbf{Rotation Policies (CRDs/Annotations):} Define frequency (rotationInterval), strategy (e.g., ``RollingUpdate''), and conditions for rotation.
    \item \textbf{Kubernetes API Integration:} All actions leverage the Kubernetes API, ensuring compliance with native security constructs \cite{kubernetes_security}.
\end{itemize}

This architecture is non-intrusive to the AI application (e.g., NIM \cite{nvidia_nim}) itself.

\subsection{Workflow: Automated and Continuous Refresh}
ADA performs rotation by deleting and recreating pods on a fixed schedule, defined by a rotationInterval. This interval is configurable per workload via Kubernetes annotations or CRDs. By default, each pod is treated as stateless and ephemeral--only the container runtime, memory state, and temporary storage are recycled. Persistent resources (e.g., volumes, config maps) remain unchanged unless explicitly mutated.

\begin{enumerate}
    \item \textbf{Onboarding:} AI workload (for example, NIM deployment \cite{nvidia_nim}) is annotated or linked to an ADA policy.
    \item \textbf{Monitoring \& Policy Evaluation:} ADA Controller tracks instance age against rotationInterval.
    \item \textbf{Rotation Execution:} The controller triggers a Kubernetes rolling update (or similar strategy) to replace old pods with new clean instances from the original image. This is seamless for stateless services like most NIMs \cite{nvidia_nim,kubernetes_security}.
    \item \textbf{Invalidation of Attacker Assumptions:} Each rotation aims to: \begin{itemize}
        \item Change network presence (new IP of the pod).
        \item Wipe in-memory attacker artifacts.
        \item Remove the exploited state specific to the instance.
        \item Nullify compromised credentials within the pod.
        \item Increase the cost of persistence of the attacker significantly \cite{mtd_survey}.
    \end{itemize}
\end{enumerate}

\section{ADA vs. A2A Security: A Leap in Simplicity and Agility}
ADA's innovative posture is clear when contrasted with traditional or complex Agent-to-Agent (A2A) security controls, which often layer on new protocols and identity management for interservice trust.

\begin{itemize}
    \item \textbf{Deployment Simplicity:} ADA uses existing Kubernetes knowledge \cite{kubernetes_security}, unlike A2A models that might require the adoption of new SDKs or protocols by application developers.
    \item \textbf{Maintenance Efficiency:} Zero-trust via rotation sidesteps complex agent identity lifecycles or A2A schema management. Security is an emerging property of the behavior of the infrastructure.
    \item \textbf{Rapid Adoption:} Familiarity for Kubernetes teams accelerates uptake.
    \item \textbf{Pragmatic Zero-Trust:} ADA offers zero trust ``by design'' for the instance itself by making it transient, reducing the reliance on intricate, potentially fallible cryptographic verification between every instance of an ephemeral agent. The security of the Kubernetes control plane itself remains paramount \cite{kubernetes_security}.
\end{itemize}

While A2A security is vital for certain interaction patterns, ADA argues that for securing the AI workload instance itself, continuous rotation offers a more robust and operationally simpler path within Kubernetes. This approach reduces the temporal attack surface in any single instance \cite{mtd_survey}.

\begin{figure}[htbp]
\centering
\includegraphics[width=0.7\textwidth]{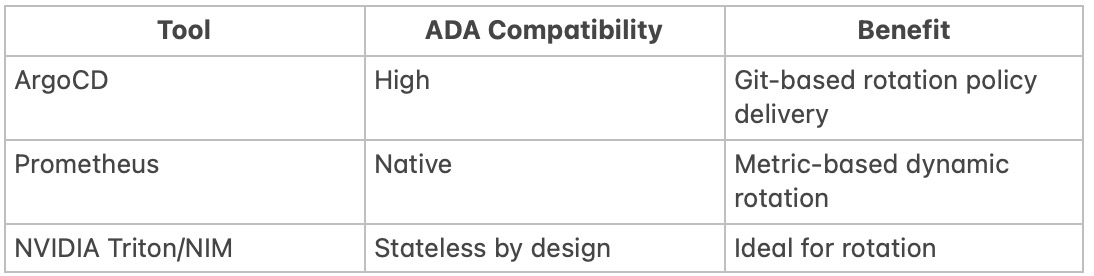}
\caption{ADA vs A2A Security Comparison Matrix}
\label{fig:ada_vs_a2a}
\end{figure}

\section{Security Considerations}
ADA may be implemented as a multi-agent system. This section outlines its security considerations using the MAESTRO framework for structured threat modeling.

MAESTRO (Multi-Agent Environment, Security, Threat, Risk, and Outcome) is a layered and architectural framework for threat modeling multi-agent systems. It structures the analysis across seven distinct layers, from foundation models to the agent ecosystem, enabling identification of layer-specific and cross-layer vulnerabilities. MAESTRO emphasizes the dynamic and interconnected nature of agentic systems, providing a comprehensive approach to security that considers AI-specific risks, traditional security principles, and the overall system architecture.

\subsection{Threat Model: ADA and MAESTRO}
ADA's core mission is to thwart attackers seeking persistence on AI workload instances (e.g., NIM pods \cite{nvidia_nim}) for data/model theft, lateral movement, or resource abuse. We can more rigorously model this threat landscape using the MAESTRO framework to better cover a range of potential vulnerabilities in our environment.

\subsubsection{MAESTRO-Guided Threat Analysis}
The following analysis maps ADA components and mitigations in the seven architectural layers of MAESTRO to provide a comprehensive threat model. Note that many of the cross-layer threats are the same as the core threats in a secure MAESTRO environment.

\paragraph{Layer 1: Foundation Models}
\begin{itemize}
    \item \textbf{Threat:} Model Poisoning (Indirect) - While ADA does not directly influence model training, a compromised container (Layer 4) could exfiltrate data used for future fine-tuning.
    \item \textbf{Mitigation:} Robust image scanning and CI/CD policies (Layer 4) to ensure trusted base images.
\end{itemize}

\paragraph{Layer 2: Data operations}
\begin{itemize}
    \item \textbf{Threat:} Data Exfiltration (Indirect) - Attackers could exfiltrate model weights or other data before rotation occurs if proper access controls are not in place (Layer 6).
    \item \textbf{Mitigation:} Fine-grained access controls (RBAC, Network Policies - Layer 6) restrict data access. Secure credential management (Layer 6) prevents credential theft.
\end{itemize}

\paragraph{Layer 3: Agent framework}
\begin{itemize}
    \item \textbf{Threat:} Privilege Escalation via CRDs - Maliciously crafted ADA CRDs or vulnerabilities in the ADA controller code itself could lead to privilege escalation within the Kubernetes cluster (Layer 4).
    \item \textbf{Mitigation:} Rigorous validation of CRDs, minimal permissions for the ADA controller service account, and code reviews (Layer 6). Protect admission controllers.
    \item \textbf{Threat:} Tool Misuse - An attacker gains knowledge of internal APIs and calls in ADA to manipulate the system.
    \item \textbf{Mitigation:} Apply least privilege principles and RBAC.
\end{itemize}

\paragraph{Layer 4: Deployment \& Infrastructure}
\begin{itemize}
    \item \textbf{Threat:} Compromised Container Images - Malicious code could be injected into the container images used by ADA, compromising workload instances.
    \item \textbf{Mitigation:} Image signing, enforced scanning, and policy setting in CI / CD pipelines. Use short-lived credentials and external secret managers.
    \item \textbf{Threat:} Orchestration Attacks - Vulnerabilities in Kubernetes itself or misconfigurations could be exploited for unauthorized access/control.
    \item \textbf{Mitigation:} Regular security audits, adherence to Kubernetes best practices, and use of tools such as OPA Gatekeeper or Kyverno for policy enforcement.
    \item \textbf{Threat:} Lack of network policies - Open network policies can expose lateral movement opportunities.
    \item \textbf{Mitigation:} Apply K8s Network Policies using a service mesh.
\end{itemize}

\paragraph{Layer 5: Evaluation \& Observability}
\begin{itemize}
    \item \textbf{Threat:} Insufficient Logging/Visibility - If ADA rotations aren't properly logged and monitored, detecting persistent threats becomes difficult.
    \item \textbf{Mitigation:} Integrate ADA with observability tools (Prometheus, Grafana) to track rotation frequency, failures, and other key metrics. Ensure proper audit trails. Limit automated changes and keep audit trails to monitor the state.
    \item \textbf{Threat:} Over-rotation covering persistent threats: If ADA's rotation interval is too fast, then a persistent threat could be covered up.
    \item \textbf{Mitigation:} Integrate observability tools and security auditing to ensure that this is not the case.
\end{itemize}

\paragraph{Layer 6: Security and compliance}
\begin{itemize}
    \item \textbf{Threat:} Poor RBAC Settings - Overly permissive RBAC settings for the ADA controller service account could lead to privilege escalation.
    \item \textbf{Mitigation:} Enforce least privilege for the ADA controller. Implement data plane or workload-level data auditing (Falco).
    \item \textbf{Threat:} Evasion - Attackers use adversarial techniques to bypass the system.
    \item \textbf{Mitigation:} Data Poisoning AI driven input validation; rerandomization of agent containers.
\end{itemize}

\paragraph{Layer 7: Agent ecosystem}
\begin{itemize}
    \item \textbf{Threat:} Overly permissive deployment policies: enforce minimum privilege.
    \item \textbf{Mitigation:} Validate deployments policies using K8s Network Policies.
\end{itemize}

\subsubsection{Cross-Layer Threats}
\begin{itemize}
    \item \textbf{Privilege Escalation \& Lateral Movement (Layers 4, 6, 7):} An attacker gaining initial access to a container could exploit overly permissive RBAC (Layer 6) and network policies (Layer 7) to move laterally within the cluster and eventually compromise the Kubernetes control plane (Layer 4).
    \item \textbf{Supply Chain Compromise (Layers 1, 3, 4):} A tainted base image (Layer 1) could be used in the container image (Layer 4), which in turn affects the ADA controller and thus affects everything.
    \item \textbf{Over-Rotation Obscuring Threats (Layers 4, 5):} Overfrequent rotation without proper monitoring (layer 5) could mask persistent threats. An attacker might gain a foothold, but the rapid rotation cycle simply wipes them out before they can be detected, allowing them to reinfect a fresh instance shortly after.
\end{itemize}

By systematically analyzing each layer and the interactions between layers, we develop a more comprehensive understanding of the potential threats facing ADA. This informs the selection of effective mitigation strategies that could be added to Kubernetes-based ADA deployments.

\begin{figure}[htbp]
\centering
\includegraphics[width=0.99\textwidth]{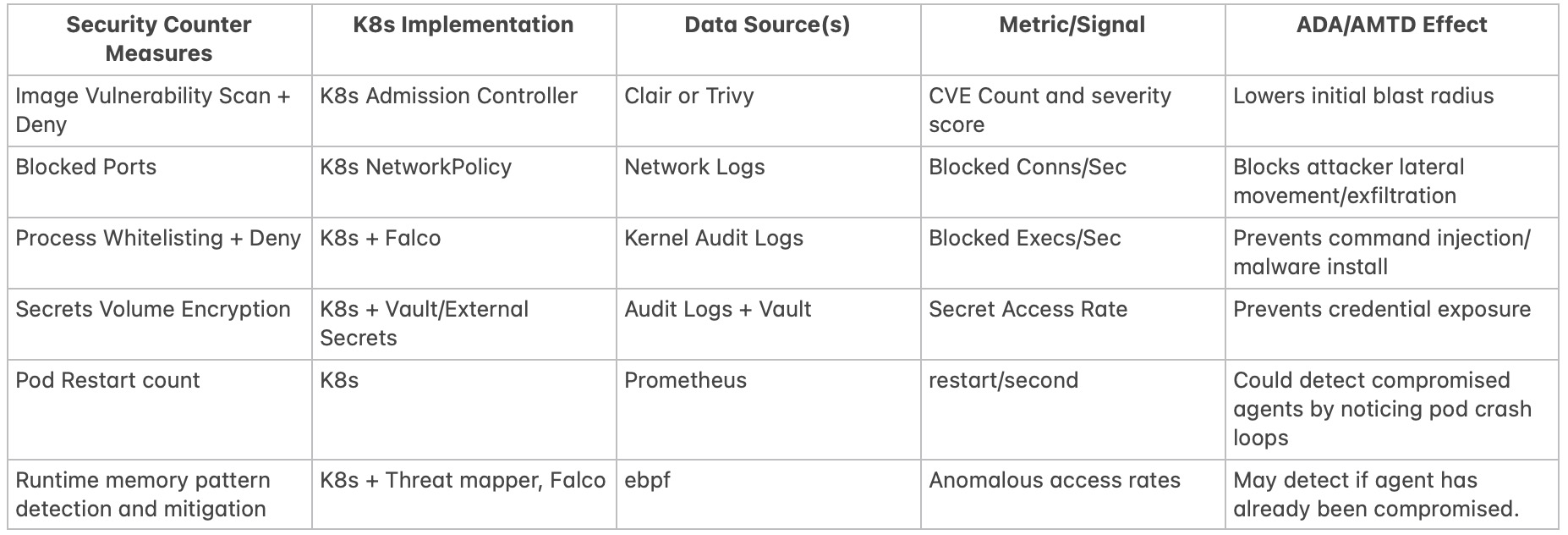}
\caption{ADA Performance Metrics and Thresholds}
\label{fig:metrics}
\end{figure}

\subsubsection{Threat Mitigation Coverage}
\begin{figure}[htbp]
\centering
\includegraphics[width=0.7\textwidth]{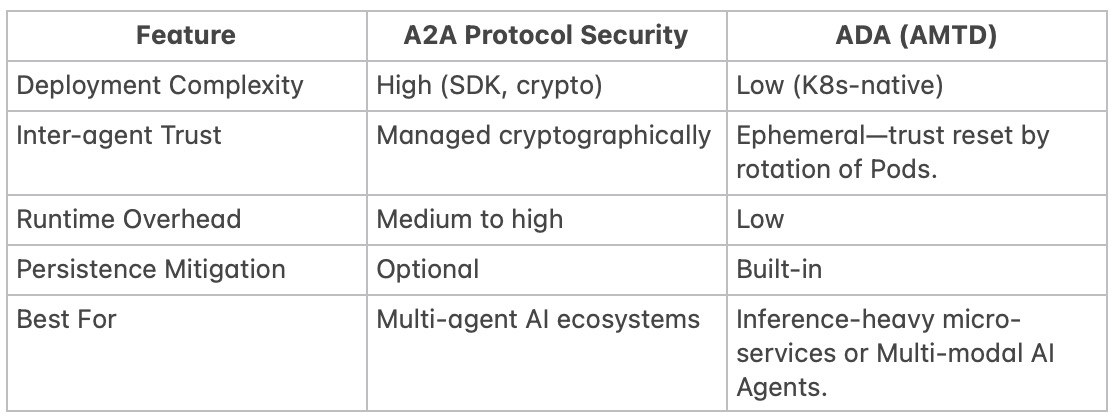}
\caption{Threat Coverage Enabled by Adaptive AMTD}
\label{fig:threat_coverage}
\end{figure}

Figure~\ref{fig:threat_coverage} outlines the effectiveness of AMTD against high value AI agent attack vectors.

The following attack vectors and their corresponding AMTD strategies demonstrate ADA's comprehensive threat mitigation approach:

\begin{itemize}
    \item \textbf{Model Exfiltration:} Distributed model sharding and randomized memory layouts
    \item \textbf{Pod Escalation:} Ephemeral pod lifecycle disrupts attacker persistence
    \item \textbf{Data Poisoning:} AI-driven input validation; agent container re-randomization
    \item \textbf{Lateral Movement:} Rotated namespaces and dynamic NetworkPolicy enforcement
    \item \textbf{Prompt Injection \& Replay:} Agent runtime environment mutated per-session
\end{itemize}

\paragraph{Integration with Agentic AI Platforms}
Agentic workloads based on JSON-RPC or LangGraph routing benefit from AMTD in the following ways:

\begin{itemize}
    \item \textbf{Adaptive Routing Control:} ADA-AMTD dynamically reroutes agent invocations via updated paths in the orchestration DAG upon anomaly detection.
    \item \textbf{Runtime RAG Sandbox:} Retrieval-Augmented Generation (RAG) components run in isolated pods with limited time-bound data exposure.
    \item \textbf{Short-Lived Checkpointing:} Stateful agent memory is written to encrypted ephemeral volumes, which are rotated and re-keyed on every inference cycle.
\end{itemize}

\paragraph{Operational Considerations}
To ensure low disruption, AMTD's rotation cadence is adjusted based on service health metrics (latency, throughput) and threat risk scores. Mutation policies are versioned via GitOps pipelines (e.g., ArgoCD), and observability integrations filter out expected noise from benign mutations.

\subsection{Metrics}
To validate the effectiveness of ADA, we propose monitoring the following key metrics:

\begin{itemize}
    \item \textbf{Time-to-Evict (TTE):} Should align with rotationInterval.
    \item \textbf{Attacker Effort Increase:} Quantified by disruption to kill chains.
    \item \textbf{Service Impact:} Measured by uptime and latency during rotation.
    \item \textbf{Resource Overhead:} Controller and churn impact, measured against established baselines.
\end{itemize}

\subsection{Other Security Considerations}
\subsubsection{Misconfigurations}

\textbf{Threats:}
\begin{enumerate}
\item Poor RBAC settings that enable privilege escalation
\item Overly permissive policies allowing lateral movement
\item Insufficient access control boundaries
\end{enumerate}

\textbf{Mitigations:}
\begin{enumerate}
\item Enforce least privilege principles for ADA controller service accounts
\item Implement rigorous CRD validation and admission control
\item Deploy comprehensive Kubernetes network policies
\end{enumerate}

\subsubsection{Accounts \& Secrets}

\textbf{Threats:}
\begin{enumerate}
\item Hardcoded secrets embedded in container images or configuration
\item Leaked service account tokens enabling unauthorized access
\item Static credentials without rotation mechanisms
\end{enumerate}

\textbf{Mitigations:}
\begin{enumerate}
\item Implement short-lived credential mechanisms with automatic renewal
\item Integrate external secret management systems (e.g., HashiCorp Vault)
\item Establish automated secret rotation policies aligned with pod rotation intervals
\end{enumerate}

\subsubsection{Ephemeral Infrastructure}

\textbf{Threats:}
\begin{enumerate}
\item Predictable naming conventions enabling targeted attacks
\item Persistent data remnants surviving rotation cycles
\item Brief but exploitable windows during pod transition states
\end{enumerate}

\textbf{Mitigations:}
\begin{enumerate}
\item Implement cryptographically secure randomized naming schemes
\item Utilize ephemeral volumes with secure data wiping mechanisms
\item Deploy anomaly-triggered emergency rotation procedures
\end{enumerate}

\subsubsection{Supply Chain}

\textbf{Threats:}
\begin{enumerate}
\item Tainted container images containing malicious code or vulnerabilities
\item Compromised CI/CD pipelines injecting malware into build artifacts
\item Unverified third-party components with unknown security posture
\end{enumerate}

\textbf{Mitigations:}
\begin{enumerate}
\item Implement cryptographic image signing and verification workflows
\item Enforce comprehensive vulnerability scanning with policy-based gating
\item Establish secure software supply chain practices with provenance tracking
\end{enumerate}

\subsubsection{Tenancy Boundaries}

\textbf{Threats:}
\begin{enumerate}
\item Namespace collisions enabling cross-tenant data exposure
\item Unauthorized inter-tenant access through misconfigured boundaries
\item Resource exhaustion attacks affecting multiple tenants
\end{enumerate}

\textbf{Mitigations:}
\begin{enumerate}
\item Implement strict namespace scoping with cryptographic tenant isolation
\item Enforce multi-layered tenant policies with admission controllers
\item Deploy network-level workload isolation using service mesh technologies
\end{enumerate}

\subsubsection{Runtimes \& APIs}

\textbf{Threats:}
\begin{enumerate}
\item Maliciously crafted Custom Resource Definitions enabling cluster compromise
\item Unvalidated API interactions bypassing security controls
\item Privilege escalation paths through Kubernetes RBAC misconfigurations
\end{enumerate}

\textbf{Mitigations:}
\begin{enumerate}
\item Deploy robust admission controller frameworks with comprehensive validation
\item Implement strict CRD schema validation and security policy enforcement
\item Enable comprehensive API auditing with anomaly detection capabilities
\end{enumerate}

\subsubsection{Operations}

\textbf{Threats:}
\begin{enumerate}
\item Excessive rotation frequency masking persistent threat indicators
\item Insufficient observability into rotation-induced system state changes
\item Loss of forensic evidence due to rapid instance turnover
\end{enumerate}

\textbf{Mitigations:}
\begin{enumerate}
\item Integrate comprehensive observability platforms with rotation-aware telemetry
\item Implement adaptive rotation policies based on threat intelligence feeds
\item Maintain persistent audit trails with cryptographic integrity protection
\end{enumerate}

\begin{figure}[htbp]
\centering
\includegraphics[width=0.7\textwidth]{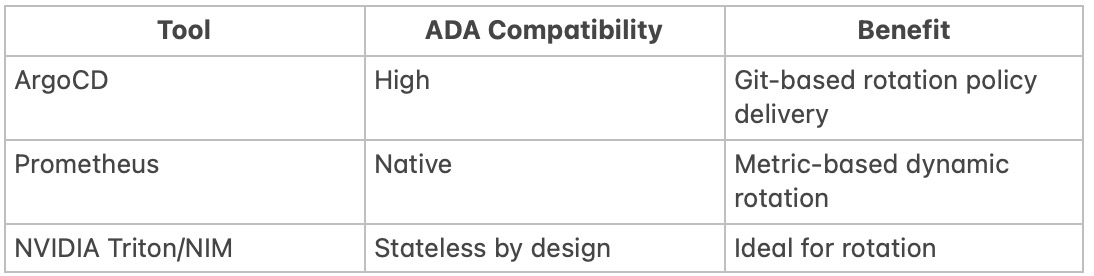}
\caption{Security Considerations and Mitigation Strategies}
\label{fig:security_considerations}
\end{figure}

\subsection{Example Scenario: Compromised NIM}
A compromised NIM instance \cite{nvidia_nim} is automatically replaced by ADA within the rotation interval, thwarting prolonged attacks. Without ADA, the compromise could persist indefinitely.

\subsection{Applicability}
The proposed approach is ideal for stateless AI inference services (NIMs \cite{nvidia_nim,a2a_security}), high-risk services, and environments with rapid CI/CD, leveraging Kubernetes \cite{kubernetes_security,aws_eks}.

\subsection{Threat Mitigation Coverage}
Table~\ref{tab:threat_coverage_detailed} outlines the effectiveness of AMTD against high value AI agent attack vectors.

\begin{table}[htbp]
\centering
\caption{Threat Coverage Enabled by Adaptive AMTD}
\label{tab:threat_coverage_detailed}
\begin{tabular}{|l|l|}
\hline
\textbf{Attack Vector} & \textbf{AMTD Strategy} \\
\hline
Model Exfiltration & Distributed model sharding and randomized memory layouts \\
\hline
Pod Escalation & Ephemeral pod lifecycle disrupts attacker persistence \\
\hline
Data Poisoning & AI-driven input validation; agent container re-randomization \\
\hline
Lateral Movement & Rotated namespaces and dynamic NetworkPolicy enforcement \\
\hline
Prompt Injection \& Replay & Agent runtime environment mutated per-session \\
\hline
\end{tabular}
\end{table}

\subsubsection{Integration with Agentic AI Platforms}
Agentic workloads based on JSON-RPC or LangGraph routing benefit from AMTD in the following ways:

\begin{itemize}
    \item \textbf{Adaptive Routing Control:} ADA-AMTD dynamically reroutes agent invocations via updated paths in the orchestration DAG upon anomaly detection.
    \item \textbf{Runtime RAG Sandbox:} Retrieval-Augmented Generation (RAG) components run in isolated pods with limited time-bound data exposure.
    \item \textbf{Short-Lived Checkpointing:} Stateful agent memory is written to encrypted ephemeral volumes, which are rotated and re-keyed on every inference cycle.
\end{itemize}

\subsubsection{Operational Considerations}
To ensure low disruption, AMTD's rotation cadence is adjusted based on service health metrics (latency, throughput) and threat risk scores. Mutation policies are versioned via GitOps pipelines (e.g., ArgoCD), and observability integrations filter out expected noise from benign mutations.

\section{Discussion: The ADA Advantage and Future}

\subsection{Benefits: A proactive and resilient defense}
ADA offers proactive security, resilience against instance-specific exploits, a drastically reduced attacker window, and operational simplicity for Kubernetes users \cite{mtd_survey,kubernetes_security}. Its innovative strength is making the environment itself the primary defense.

\subsection{Limitations}
\begin{itemize}
    \item \textbf{Stateful Applications:} The main challenge requiring complex state migration strategies.
    \item \textbf{Ultra-Fast Attacks:} May not prevent the initial ``smash and grab'' but clean up afterward.
    \item \textbf{Performance:} Churn needs careful management \cite{kubernetes_security}.
    \item \textbf{Orchestrator Security:} Relies on a secure Kubernetes control plane \cite{kubernetes_security}.
\end{itemize}

\subsection{Future Work: Expanding Innovation}

\subsubsection{Formal Security Modeling}
While ADA demonstrates a strong conceptual foundation, future work should focus on formally quantifying its security impact. This includes modeling attacker dwell time, rotation-induced uncertainty, and disruption of kill chains using techniques such as attack graphs, game theory, or simulation frameworks. Such models will provide measurable assurance of ADA's Moving Target Defense (MTD) benefits.

\subsubsection{Rotation-Aware Observability}
Frequent instance rotation can produce telemetry churn and false positives in conventional monitoring stacks. ADA should incorporate rotation-aware observability by tagging pod lifecycle events and enriching logs and metrics with rotation context. This allows tools such as Prometheus, Grafana, and SIEM systems to suppress noise and retain clarity during normal churn cycles.

\subsubsection{Adaptive Rotation Based on Risk Scores}
Rather than relying solely on time-based or static thresholds, ADA can evolve to support rotation policies based on dynamic risk signals. These could be derived from ML-based anomaly detection, policy engines (e.g., OPA), or behavioral scoring. Higher-risk workloads could be rotated more aggressively, aligning defense efforts with situational awareness.

\subsubsection{Integration with Confidential Computing}
To extend the protection to the data in use, ADA can be adapted to run within confidential computing environments such as Intel TEE, SGX or AMD SEV. By rotating pods inside secure enclaves, ADA can combine workload churn with hardware-backed isolation, addressing advanced threat scenarios, including insider attacks, lateral movement, or host compromise.

\subsubsection{Extension to Stateful Workloads}
Currently optimized for stateless services, ADA's applicability to stateful or long-running tasks (e.g., fine-tuning or reinforcement learning agents) requires future support for state checkpointing and safe rehydration strategies. Exploring rotation-aware storage, such as volume snapshotting or sidecar-based state offloading, will expand the scope of ADA to a broader class of AI workloads.

\subsubsection{Secure Sandboxed Execution for Agentic Code}
For multi-agent or GenAI workloads that may execute untrusted or self-modifying code, lightweight sandboxing tools like Pyodide (Python in WebAssembly) and Deno (secure JavaScript/TypeScript runtime) can be integrated. These frameworks provide run-time confinement and synchronous restrictions at the application layer. While not part of ADA's core, integrating sandboxing at the pod level complements rotation by adding execution isolation within each short-lived instance.

\section{Conclusion}
ADA presents a forward-thinking AMTD system that leverages the inherent dynamism of Kubernetes to provide robust security for AI workloads such as NIMs \cite{mtd_survey,nvidia_nim,kubernetes_security}. By making instances ephemeral through continuous rotation, it offers a simpler, more agile approach than complex A2A security frameworks for instance hardening. This innovative use of infrastructure as an active defense mechanism marks a significant step towards building intrinsically secure AI systems. ADA champions a shift from static fortifications to dynamic and resilient defenses, making the operational environment a core component of its own protection.

\section{Disclaimer}
The views and opinions expressed in this paper are those of the authors and do not necessarily reflect the official policy or position of their respective organizations. The research presented here is for open source, academic, research, and educational purposes only. There is no IP or copyright violation by doing this open source AI research work for creating and publishing this paper.

\bibliographystyle{ieeetr}

\end{document}